\documentclass[journal,12pt,onecolumn]{IEEEtran}
\IEEEoverridecommandlockouts
\ifCLASSINFOpdf
\else
\fi
\usepackage{graphicx}
\usepackage{latexsym,bm}
\usepackage{amsmath}
\usepackage{amssymb}
\usepackage{amsthm}
\usepackage{epstopdf}
\usepackage{cite}
\usepackage{amsfonts}
\usepackage{mathrsfs}
\usepackage{color}
\newcommand\blfootnote[1]{%
  \begingroup
  \renewcommand\thefootnote{}\footnote{#1}%
  \addtocounter{footnote}{-1}%
  \endgroup
}
\begin{document}
\begin{titlepage}
\begin{center}
\vspace*{-2\baselineskip}
\begin{minipage}[l]{7cm}
\flushleft
\includegraphics[width=2 in]{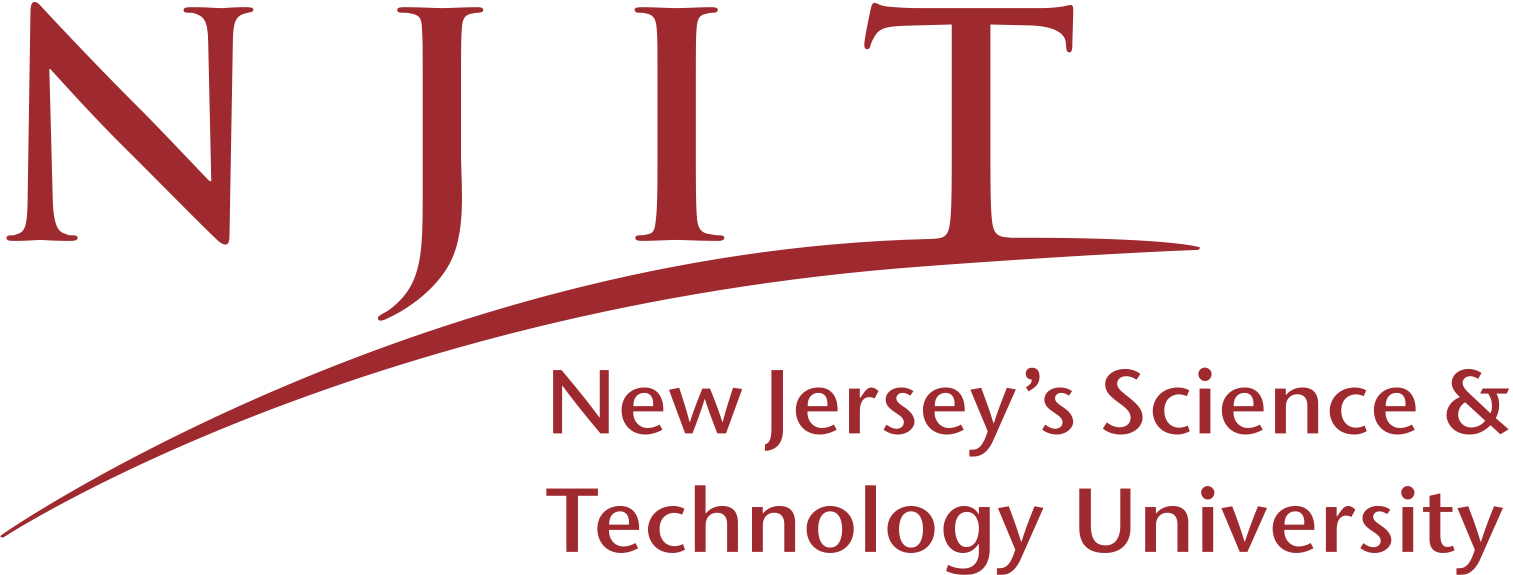}
\end{minipage}
\hfill
\begin{minipage}[r]{7cm}
\flushright
\includegraphics[width=1 in]{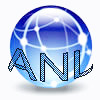}%
\end{minipage}

\vfill

\textsc{\LARGE On Green Energy Powered \\ [12pt]
Cognitive Radio Networks}

\vfill
\textsc{\LARGE XUEQING HUANG\\[12pt]
TAO HAN\\[12pt]
\LARGE NIRWAN ANSARI}\\
\vfill
\textsc{\LARGE TR-ANL-2014-003\\[12pt]
\LARGE May 22, 2014}\\[1.5cm]
\vfill
{ADVANCED NETWORKING LABORATORY\\
 DEPARTMENT OF ELECTRICAL AND COMPUTER ENGINEERING\\
 NEW JERSY INSTITUTE OF TECHNOLOGY}
\end{center}
\end{titlepage}


\begin{abstract}
Green energy powered cognitive radio (CR) network is capable of liberating the wireless access networks from spectral and energy constraints. The limitation of the spectrum is alleviated by exploiting cognitive networking in which wireless nodes sense and utilize the spare spectrum for data communications, while dependence on the traditional unsustainable energy is assuaged by adopting energy harvesting (EH) through which green energy can be harnessed to power wireless networks. Green energy powered CR increases the network availability and thus extends emerging network applications. Designing green CR networks is challenging. It requires not only the optimization of dynamic spectrum access but also the optimal utilization of green energy. This paper surveys the energy efficient cognitive radio techniques and the optimization of green energy powered wireless networks. Existing works on energy aware spectrum sensing, management, and sharing are investigated in detail. The state of the art of the energy efficient CR based wireless access network is discussed in various aspects such as relay and cooperative radio and small cells. Envisioning green energy as an important energy resource in the future, network performance highly depends on the dynamics of the available spectrum and green energy. As compared with the traditional energy source, the arrival rate of green energy, which highly depends on the environment of the energy harvesters, is rather random and intermittent. To optimize and adapt the usage of green energy according to the opportunistic spectrum availability, we discuss research challenges in designing cognitive radio networks which are powered by energy harvesters.
\end{abstract}
\blfootnote{This work was supported in part by NSF under grant no. CNS-1320468.}

\section{Introduction}
Wireless access networks are among the major energy guzzlers of the telecommunications infrastructure, and their current rate of power consumption is escalating because of the explosive surge of mobile data traffic \cite{6472203}, \cite{DeDomenico20145}. Therefore, greening wireless access networks will not only decrease the operational expenditures (OPEX), but also improve the sustainability of information communications technology (ICT).

Greening wireless access networks can capitalize on the broad paradigm of cognitive radio (CR). Haykin \cite{Def} has precisely defined CR as ``A cognitive radio transmitter will learn from the environment and adapt its internal states to statistical variations in the existing radio frequency (RF) stimuli by adjusting the transmission parameters (e.g., frequency band, modulation mode, and transmission power) in real-time and on-line manner". With the \emph{cognitive capability} to detect the available spectrum and the \emph{reconfigurability} to dynamically access spectrum over which less fading and interference is experienced, the intelligent CR communication system enhances the spectrum agility and energy efficiency \cite{disertation}.

Among many areas of wireless systems that can be improved via CR, existing literature primary focuses on improving spectrum efficiency rather than energy utilization, let alone the unreliable dynamic energy. Zhao and Sadler \cite{threestrategies} presented an overview of the dynamic spectrum access (DSA), which differentiates the CR wireless system from legacy wireless systems, which are licensed to operate in a dedicated operating frequency band. Various functionalities of CR are introduced in \cite{Func}, \cite{SM}. While researchers in 
\cite{SSsingleDualRadio} - \cite{5703204} discussed spectrum sensing, which enables CR to seek and access the spectrum opportunities, Wang \emph{et al.} \cite{5723803} surveyed various applications which can take advantages of the DSA feature of CR, and Akyildiz \emph{et al.} \cite{HO} discussed the CR-based wireless access network architecture.
 
Different from current CR networks powered by the reliable on-grid or un-rechargeable energy source, continuous advances in green energy have motivated us to focus on the energy efficiency and investigate the \emph{green powered cognitive radio (green CR) network}. The concept of energy harvester has been proposed to capture and store ambient energy to generate electricity or other energy form, which is renewable and more environmentally friendly than that derived from fossil fuels \cite{RF}, \cite{RFHARVESTOR}. If the green energy source is ample and stable in the sense of availability, the cognitive radio network can be powered to opportunistically exploit the underutilized spectrum by harnessing free energy without requiring energy supplement from external power grid or battery.

It is, however, not trivial to design and optimize the green energy enabled CR networks owing to the opportunistic arrival of idle spectrum and free energy in reality. This survey concisely summarizes the state-of-the-art research in energy efficient cognitive radio systems from three aspects: 1) achieving power aware functionality in the CR systems, 2) designing energy efficient wireless access systems via cognitive radio, and 3) optimizing green CR networks. Although green powered wireless networks currently are not deployed at a large scale owing to the higher cost per watt than the grid energy, powering wireless networks by green energy is imminent and is becoming a sustainable and economically friendly solution. The aim of this article is to provide some insights for future research in green powered dynamic spectrum access-based cognitive radio networks, which are capable of liberating wireless access networks from spectral and energy constraints. 

The rest of the paper is organized as follows. After reviewing the fundamental concepts of cognitive radio and dynamic spectrum access, we provide 1) a summary of energy efficient spectrum sensing, spectrum management and handoff, and spectrum sharing methods for cognitive radio systems in Section II. This is followed by 2) the current and future trends of cognitive radio based wireless access networks, e.g., energy efficient relay/cooperation cognitive radio system and cognitive small cells, in Section III. Then, 3) the energy-harvesting based cognitive radio functionality is covered in Section IV. To this end, 4) the optimization of green energy utilization in cognitive radio networks (CRNs) are reviewed and open research issues are outlined in Section V. 
\begin{figure*}
\includegraphics[width=\textwidth]{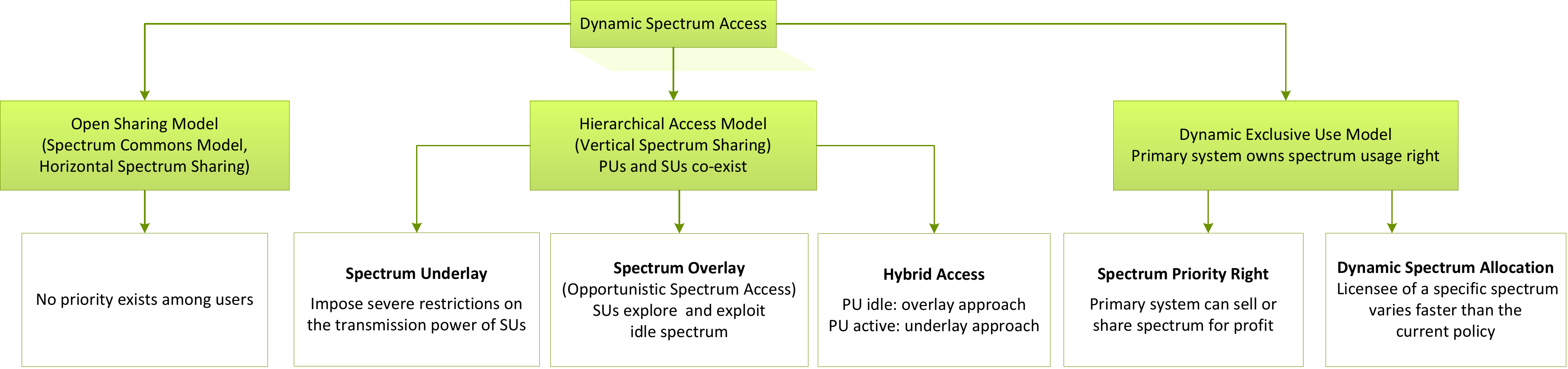}
\caption{A taxonomy of dynamic spectrum access (expanded and adopted from \cite{threestrategies},\cite{4221550})}
\label{taxonomy}
\end{figure*}

\section{Energy Efficiency in CR}
Traditional spectrum licensing schemes use the command-and-control model, in which explicit rules are laid for use of the
spectrum \cite{4221550}. Severe under-utilization of the static licensed spectrum presents great challenges for the resource constrained wireless network to handle the growing popularity of new wireless devices and applications. To enable more flexible spectrum access, dynamic spectrum access techniques have been proposed to solve this spectrum inefficiency problem, by allowing unlicensed users to access the radio spectrum under certain restrictions.


In a dynamic spectrum access network, three broadly categorized models (based on \cite{threestrategies},\cite{4221550}) shown in Fig. \ref{taxonomy} are commonly used. Cognitive radio technology has emerged as a key enabler for dynamic spectrum access. In a CR-based DSA network, the primary system owns the spectrum rights, while unlicensed users can dynamically share the licensed spectrum in an opportunistic manner. This capability is provided by the following cognitive functionalities \cite{Func}: 1) \emph{Spectrum sensing and analysis}: Spectrum sensing is defined as the task of finding spectrum opportunities, i.e., spectrum holes, in the local neighborhood of the overlay CR receiver \cite{SS}. Once the spectrum hole is utilized by a secondary (unlicensed) transmitter, no primary receiver will be affected by this secondary transmitter, and no primary transmitter will interfere with the intended secondary receiver either \cite{specrumHole1}. Spectrum analysis uses the information obtained from spectrum sensing to schedule and make a decision to access the spectrum by SUs. 2) \emph{Spectrum management and handoff}: This function enables SUs to select the best available channel and vacate the channel when PU is detected to reclaim it \cite{HO}, and therefore somewhat addresses the challenges imposed by the fluctuating nature of the available spectrum, as well as the diverse quality of service (QoS) requirements of various applications \cite{SM}. 3) \emph{Spectrum allocation and sharing}: It coordinates access to the available channel with other users (PUs, other SUs, or both), so that the interference level incurred by secondary spectrum usage should be kept below a certain threshold, and collisions and interference among multiple SUs are alleviated.

Although CRs enable spectrum sharing with smart operation and agile spectrum access \cite{greenCRINTRODUCTION}, a major limitation of a practical cognitive radio network is the increasing power consumption introduced by the cognitive capability and reconfigurability. Energy aware cognitive radio system has been investigated from three general perspectives. 1) \emph{Energy minimization} minimizes the power consumption for a given performance requirements. 2) \emph{Performance maximization} maximizes the performance for a given limited power budget. 3) \emph{Utility maximization} takes the power consumption cost into account during sensing/transmission, with utility functions in the form of, such as, the difference between the two, or the ratio of performance reward and cost of power.
\begin{subequations}
\label{utility_f}
\begin{align}
{{\eta}=F(P)-{\mu}P} \\
{{\eta}=\frac{F(P)}{{\mu}P}}
\end{align}
\end{subequations}
where specific performance criteria of utility $\eta$, such as 'bit/Joule', can be found in \cite{DeDomenico20145}, $F$ is the performance function with respect to power consumption $P$, and $\mu$ is a price parameter. 

Based on the various approaches and criteria mentioned above, we will introduce the functionalities of energy efficient CR.

\subsection{Spectrum Sensing and Analysis}
\begin{figure*}
\includegraphics[width=\textwidth]{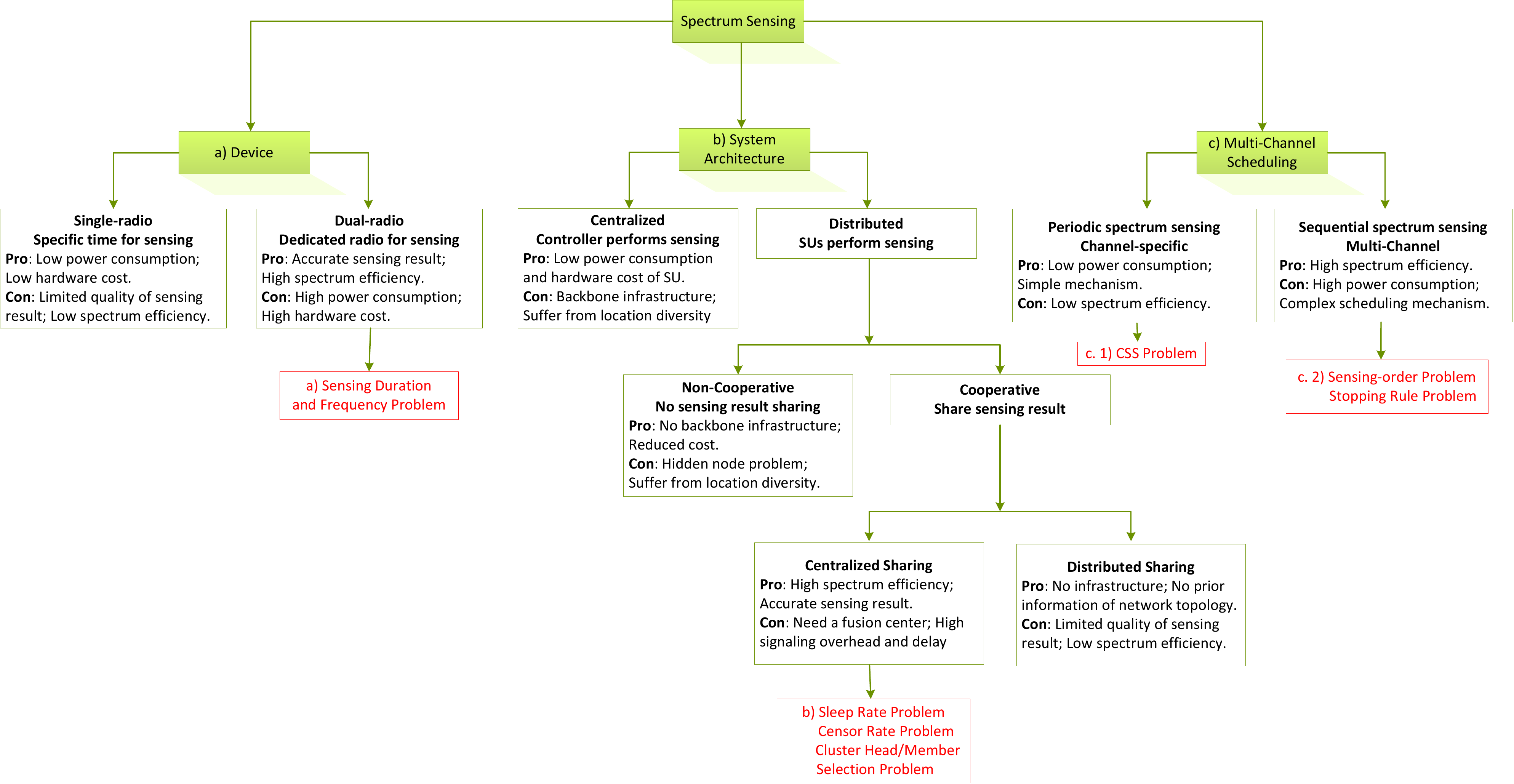}
\caption{Categories of spectrum sensing.}
\label{ss_fig}
\end{figure*}

The primary goal of CR is to access the spectrum opportunistically. This makes spectrum sensing crucial because it enables users to detect the activities of a primary user in a frequency band. The categories of spectrum sensing, energy-related advantages and drawbacks, and the research problems concerning energy efficiency are presented in Fig. \ref{ss_fig}.

To design energy efficient spectrum sensing methods, we should first learn the performance indicators used to measure the detection performance: the \emph{detection probability} $p_d$, i.e., the probability of declaring the presence of PU when it is in fact occupying the spectrum, and the \emph{false alarm probability} $p_f$, i.e., the probability of declaring the presence of PU when it is actually idle. The higher the $p_d$, the better the PUs are protected. The lower the $p_f$, the more spectrum opportunities are explored by SUs \cite{ssCriteria}. Note that the complementary probability of $p_d$, is called \emph{miss-detection probability}, and sometimes referred to as \emph{collision probability} when SU starts transmission immediately once the sensing result indicates the channel is idle \cite{threestrategies}.

To improve the quality of spectrum sensing under various conditions, such as whether \emph{prior} information about the probability of the primary user's presence on the target channel, defined in Eq. (\ref{prio_primary}), is available or not, and the target channel is narrow or wide, a collection of spectrum sensing methods have been proposed \cite{SSsingleDualRadio}.
\begin{equation}
\begin{array}{l}
{\bf{\mathscr{{\pi}}}}_0={Prob}\{\bf{\mathscr{{H}}_0}\}\\
{\bf{\mathscr{{\pi}}}}_1={Prob}\{\bf{\mathscr{{H}}_1}\}
\end{array}
\label{prio_primary}
\end{equation}
where $\bf{\mathscr{{H}}_1}$ is the hypothesis that PU is active, while $\bf{\mathscr{{H}}_0}$ means the spectrum is idle.

Although the aforementioned algorithms yield different sensing performance and power consumption, the relationship in Eq. (\ref{sense_f}) holds in general, based on the binary hypothesis testing.
\begin{equation}
\begin{array}{l}
{p_d}={Prob}\{f(T_S,{\gamma})>{{\lambda}_d}|\bf{\mathscr{{H}}_1}\}\\
{p_f}={Prob}\{f(T_S,{\gamma})>{{\lambda}_f}|\bf{\mathscr{{H}}_0}\}
\end{array}
\label{sense_f}
\end{equation}
where $f$ is related to signal detection and processing in the physical layer (PHY) \cite{SSsingleDualRadio}, \cite{HO}. $p_d$ is increasing with the sensing time $T_S$ and SNR $\gamma$, but is decreasing with the detection threshold ${\lambda}_d$. $p_f$ is decreasing with $T_S$ and $\gamma$ as well as the false alarm threshold ${\lambda}_f$.

In addition to sensing performance, the energy efficiency of the spectrum sensing algorithms depends on the power consumption, as shown in Eq. (\ref{sense_power}), which involves the scheduling of a spectrum sensing activity in the time and spatial domains by the media access control (MAC) layer.
\begin{equation}
\begin{array}{l}
{P_S}={F_S}({T_S})={F_S}({N_S})\\
{P_{R}}={F_R}({N_C,D})\\
\end{array}
\label{sense_power}
\end{equation}
where ${F_S}$ and ${F_R}$ are the sensing and reporting power functions, respectively. The sensing power $P_S$ is increasing with the sensing time $T_S$, which is further proportional to $N_S$, the number of collected samples. $P_{R}$, the power consumed by reporting the sensing result, is increasing with the transmission distance $D$ and the number of nodes participating in the sensing $N_C$. 

Besides the energy consumed to obtain the sensing result, the major energy waste of CRNs is due to the miss-detection of PU, which can lead to collision and retransmission. As a result, the power consumed for data transmission, $P_T$, is related to the detection probability ${p_d}$ and SNR $\gamma$, according to a decreasing function ${F_T}$ defined as follows.
\begin{equation}
{P_{T}}={F_T}({\gamma},{p_d})
\label{tr_power}
\end{equation}

\subsubsection{Sensing Duration and Frequency Problem} As shown in in Fig. \ref{ss_fig}a), to balance the energy-inefficiency of continuous spectrum sensing (high $P_S$) and the unreliability issue of periodic spectrum sensing (low ${p_d}$ and high ${p_f}$), Gan \emph{et al.} \cite{6253067} optimized the sampling rate (${N_S}/{T_S}$) and sensing time $T_S$ to minimize the total sensing power across multiple potential channels, within the constraints of detection performance. 

\subsubsection{Sensing Architecture Design Problem} As shown in Fig. \ref{ss_fig}b), with multiple SUs sensing and sharing information, power consumed by sensing $P_S$ and reporting $P_{R}$ can both be decreased when every cognitive radio will randomly turn off its sensing device, this energy saving approach is referred to as {\bf{\emph{sleeping}}} or on-off sensing. The probability of switching off is called the \emph{sleeping rate}. The second approach to reduce $P_{R}$ is {\bf{\emph{censoring}}}, where a sensing result is sent only if it is deemed informative. The \emph{censor rate} refers to the probability of the sensing results being located in the censor region, a signal interval for the locally collected
energy \cite{spectrumSensingSurvey1}. Another method of reducing $P_{R}$ is {\bf{\emph{clustering}}}, instead of sending local sensing results directly to the fusion center (FC), they are sent to the assigned \emph{cluster heads} (CHs), which then make local cluster decisions and send them to FC. This way the network energy consumption is reduced due to the distance reduction. 

Extensive centralized detection performance optimization and overhead minimization algorithms can be found in \cite{6036931} - \cite{DSS}, where three approaches are implemented in a complementary way. Besides the centralized spatial diversity scheduling, each SU can decide whether to sleep or share results by itself. However, the broadcasting nature of wireless environment provides the opportunity for SU to be a free rider, who overhears the sensing result and does not contribute to the sensing process. 
Various games have been modeled in \cite{5426522}, \cite{6503328} to study the above mentioned incentive issue of cooperative spectrum sensing where each selfish SU aims to maximize its own utility in the form of Eq. (\ref{utility_f}a).


\subsubsection{Multi-Channel Scheduling Problem} With the introduction of scheduling in the frequency domain, spectrum opportunity can be further exploited from a new degree of freedom, as illustrated in Fig. \ref{ss_fig}c).

\emph{3.1)} {Cooperative Sensing Scheduling:} When periodic spectrum sensing is adopted for multiple channels, the sensing scheduling has to address the cooperative sensing scheduling (CSS) problem \cite{cooperativeSS}: how to assign secondary nodes to different channels? Except for the \emph{energy-performance tradeoff} (greater $N_C$ leads to a better sensing performance but more energy will be consumed in both sensing and reporting), \emph{performance-opportunity tradeoff} exists in CSS, i.e., assigning more SUs to one channel will yield more reliable sensing outcome, but spectrum opportunity is not fully explored since less potential channels are sensed \cite{traDEOFFss}.

\emph{3.2)} {Sequential Channel Scheduling:} As far as the sensing process is concerned, the only task for periodic spectrum sensing is to find the idle spectrum. After getting the sensing result, SU could either transmit when the sensed channel is free, or wait for the channel which is currently occupied by PU. In sequential channel sensing, the task is to find an idle channel with good quality. So, SU can switch to a different channel and perform sensing again even when the current sensed channel turns out to be idle. A channel-specific access decision has to be made based on the analysis of both the sensing result and possible channel quality estimation afterwards if the channel is unoccupied \cite{4786510}. 

Since channel switching and sensing require additional time and energy, SU needs to efficiently locate an idle channel with satisfactory channel quality. More specifically, a \emph{channel sensing order scheme} and a \emph{channel selection protocol} (stopping rule) have to be designed to optimize CR networks.

Any performance metrics of sequential sensing scheduling will be related to the transmission performances, such as throughput and delay, since they are the criteria to judge whether the channel is good enough for SU. The transmission performance related design, which is the function of spectrum analysis and access process, will be further discussed later in the following section.

\subsection{Spectrum Management and Handoff}
Spectrum analysis and spectrum access consist of two major components of the spectrum management mechanism \cite{HO}. First, the \emph{sensing strategy} specifies whether to sense and where in the spectrum to sense. Then, spectrum analysis will estimate the characteristics of the spectrum holes that are detected through spectrum sensing. Third, the \emph{access strategy} will make a decision on whether to access or not, by judging whether this is the best available channel. While the focus of the last section (except for sequential channel scheduling) is finding the spectrum opportunity in an energy efficient way, this section will further concentrate on locating the spectrum opportunity with good quality. The channel characteristics (mostly channel capacity) determines the data transmission power level appropriate to the user requirements. As a result, energy efficiency plays an important role in the design of the spectrum management mechanism.

\subsubsection{Spectrum Access}
The sequential decision process of sensing and access strategies gives rise to a new option for SU: keeping idle on the current channel without sensing or transmission. The reason is that when energy consumption is considered, instead of sensing a channel that may turn out to be in use by PU or obtaining an idle channel with poor fading condition, it is better for SU to do nothing at all.


The partially observable Markov decision process (POMDP) framework, which permits the uncertainty of information in modeling the operation mode selection problem, is suitable for CRNs where PUs' traffic is not fully observable to the secondary network. By jointly considering three conflicting objectives among gaining immediate access, gaining spectrum occupancy information, and conserving energy (for future sensing, transmission, and idling) on a single-channel, the POMDP framework \cite{4663925} indicates that the optimal sensing decision and access strategy uses a threshold structure in terms of channel free probability and channel fading condition, i.e., SU with un-rechargeable battery and limited power should sense the channel when the conditional probability that the channel is idle in the current slot is above a certain threshold, and it will access the channel if the fading condition of this idle channel is better than a certain threshold.


\subsubsection{Spectrum Mobility}
As mentioned in the previous section, spectrum availability in CR networks varies over time and space. So, except for the traditional mobility which refers to mobile users traversing across cells or current channel conditions becoming worse \cite{doubleHandoff}, CR networks incur another dimension mobility, spectrum mobility, in which SU has to move from one spectrum hole to another to avoid interference in case of the reappearance of PU. CR networks need to perform mobility management adaptively depending on the heterogeneous spectrum availability, which is a result of the highly unpredictable mixed multimedia primary traffic in today's wireless networks.

In general, spectrum handoff, in which SU can switch to another vacant channel and continue data transmission when the current channel is sensed as busy, can be categorized into two major types. One is the \emph{reactive-sensing spectrum handoff}, in which the target channel is sensed or selected only after the spectrum handoff request is made. The other is the \emph{proactive-sensing spectrum handoff}, where the target channel for spectrum handoff is predetermined \cite{proANDreHO}. The reactive spectrum handoff pays the cost of sensing time to guarantee the accuracy of the selected target channel. By pre-determining the target channel, the proactive spectrum handoff avoids the sensing time but loses certain accuracy. Extensive performance analysis of spectrum handoff can be found in 
\cite{proANDreHO} - \cite{5285171}. 

As shown in Eq. (\ref{handoff_power}), tuning the radio frequency of the cognitive device will result in additional power consumption. 
\begin{equation}
{P_{HO}}={F_{HO}}(f_t-f_c)
\label{handoff_power}
\end{equation}
where $F_{HO}$ is an increasing function and the handing off power $P_{HO}$ is related to the frequency difference between the target channel $f_t$ and the current channel $f_c$.

In practice, to minimize the energy consumption incurred during handoff, SU may prefer to wait for the current channel and conserve energy, at the cost of service quality. As a result, the handoff strategy will have to make \emph{a priori} decision about whether to request handoff or not.



This section mainly focuses on the coordinated spectrum access among PUs and SUs, which is primarily based on the sensing result. How to coordinate spectrum access among SUs, and how to share the spectrum resource effectively when the PU is active, will be investigated in the next section.

\subsection{Spectrum Sharing and Allocation}
According to the coordination behavior between PUs and SUs, existing solutions for spectrum sharing can be classified into three major categories \cite{temANDspatialSSHARING}. 1) Spatial spectrum sharing, e.g., spectrum underlay, always allows secondary users to access the spectrum subject to \emph{interference temperature} (IT) constraint, which is the threshold of interference impairing the PU receiver \cite{interTempeSSharing}. 2) Temporal spectrum sharing, e.g., spectrum overlay, allows secondary users to utilize the spectrum only when it is idle. 3) Hybrid spectrum sharing, in which SUs initially sense for the status (active/idle) of a frequency band (as in the temporal spectrum sharing) and adapt their transmit power based on the decision made by spectrum sensing, to avoid causing heavy interference (as in spatial spectrum sharing) \cite{twoSharing}.

Since SUs might be competing for the resource or cooperating to improve efficiency and fairness of resource sharing \cite{relayDef}, \cite{spectrumSharingDefAmongSUs}, \cite{spectrumTradingDef}, spectrum sharing can be implemented in two manners: cooperative and non-cooperative, depending on the SUs' behavior within the secondary network.

Regardless of the spectrum sharing model, spectrum allocation mechanisms have a great impact on the energy consumption and performance of each individual SU and the whole secondary network. As a result, energy efficient designs, which take into account of the diversity of SUs' power budgets, the channel conditions and the QoS requirements, are of great interest to the spectrum sharing and allocation algorithms.

\subsubsection{Spatial Spectrum Sharing}
Since IT is subject to space-time-frequency variation, spectrum sharing, along with power control, bit-rate and antenna beam allocation, which are all implemented by the dynamic resource allocation (DRA), have become essential techniques for energy efficient underlay CRs \cite{5447050}. Moreover, to access the same spectrum being used by PUs, SUs normally provide incentive to compensate for the additional interference induced to PUs. The incentive can be in the form of \emph{spectrum trading}, in which radio resources such as spectrum in a CR environment can be sold and bought \cite{spectrumTradingDef}. So, efficient spectrum sharing and allocation can generate more revenue for the spectrum owner and also enhance the satisfaction of PUs.

Illanko \emph{et al.} \cite{ofdmPCnoCARRIER} considered the centralized cooperative DRA problem, where the downlink power allocation of OFDM CRN is modeled as a concave fractional program. Near optimal solutions, which maximize bit/Joule energy efficiency while meeting SUs' rate demands and PUs' IT-constraint, are given in closed form. 
In addition to the \emph{centralized DRA}, which requires a centralized entity having full knowledge of the network, \emph{distributed DRA} is more suitable for scenarios with no such infrastructure \cite{ULunderlay}. 
From PUs' viewpoint, Han and Ansari \cite{taoHan} studied the spectrum trading between the primary and secondary network. PUs' traffic is offloaded to the secondary network, which in exchange will gain spectrum for its own traffic transmission. To minimize the primary energy consumption with power budget and rate requirement, the auction-based energy-spectrum trading scheme is proposed to approximate the behavior of the primary and secondary network for distributed spectrum allocation.

\subsubsection{Temporal Spectrum Sharing}
In temporal spectrum sharing, the issue is transmission opportunity exploitation, i.e., \emph{spectrum exploitation}, which means that after a spectral hole is identified, i.e., \emph{spectrum exploration}, how efficient can SUs access and utilize the spectrum, while not interfering with the primary users. Because of the spectrum sensing process, the IT-constraint is less stringent as compared to the spatial spectrum sharing model. However, interference still exists due to frequency reuse in today's wireless systems and power leakage into the adjacent frequency band being occupied by PU \cite{5351726}. Although spectrum sensing is up to SUs, varying spectrum condition and diverse QoS requirements still challenge the energy-aware DSA.

Hasan \emph{et al.} \cite{5351726} considered the centralized power allocation over multiple OFDM subcarriers, with IT-constraint and limited transmission power budget. Sub-optimal schemes are proposed to maximize the corresponding convex utility function which incorporates the achievable data rate of SU and the expected rate loss due to sensing error or spectrum reoccupation by PU. 
Gao \emph{et al.} \cite{5288955}, on the other hand, proposed a framework of distributed DRA for energy constrained OFDMA based CRN.

 \section{Energy Efficiency via CR}

In this section, we discuss how to adopt CR into the wireless access network to improve the energy efficiency. Given the complexity of the topic and the diversity of existing technical approaches \cite{5723803}, we focus on the major applications of cooperative and cognitive radio communications, and heterogeneous cognitive radios for the emerging Long Term Evolution-Advanced (LTE-Advanced) networks. 
 
\subsection{Green Relaying and Cooperative CR}
The major challenge of CR networks is to guarantee the QoS requirements while not causing unacceptable performance degradation of PUs. In some cases, reliable end-to-end transmission within the secondary network requires a large amount of power, leading to harmful interference to PUs \cite{relayCR1}. 
The key enabling technology of boosting the overall performance while saving energy is the relay technology \cite{relayDef}. 

For the relay based CR networks, the path loss is less due to the shorter transmission range, and the interference caused to the primary network is potentially reduced due to the lower transmission power. Furthermore, when no common available spectrum between a pair of secondary users exists, the relay node can help establish the end-to-end communication through the \emph{dual-hop channel} \cite{5484603} (two sets of available spectrum: one for the source-relay link and one for the relay-destination link).

Unlike the pure relay systems, the cooperative node acts as both an information source and a relay. Network coding based two-way relay schemes with decoding (decode-and-forward) and without decoding (amplify-and-forward, denoise-and-forward, compress-and-forward) are introduced to implement cooperative communications \cite{relayDef}. The inherent cooperative diversity can save energy by combining the signals received from different spatial paths and consecutive time slots \cite{ccDEF}.

\begin{figure*}
\includegraphics[width=\textwidth]{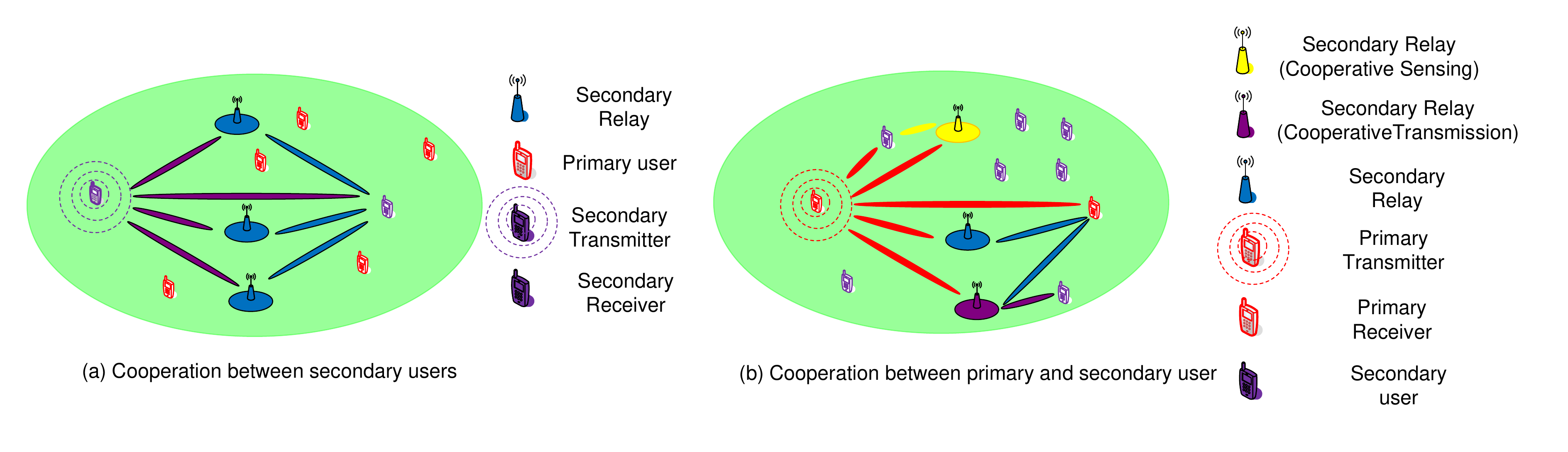}
\caption{Relaying and cooperation in cognitive radio systems.}
\label{RFandCoop_fig}
\end{figure*}

In the context of CRN, two basic cooperative transmission scenarios exist: 1) {\emph{cooperative transmission between secondary users}} that aims to increase the secondary throughput for a given spectral hole; 2) {\emph{cooperative transmission between primary and secondary users}} that aims to increase the spectrum opportunities of SU. Given the fact that under the specification of relay protocols, relay nodes will work differently from the regular cognitive devices, as illustrated in Fig. \ref{RFandCoop_fig}. The spectrum sensing and resource allocation designs need to take into account of the features of relaying cognitive radio and cooperative cognitive radio systems.


\subsubsection{Relay-enhanced Cooperative Spectrum Sensing} With the secondary relay enhanced CR networks, besides the fact that cooperation can be used to enhance the transmission performance, the capability of sensing can be improved when relays perform spectrum sensing cooperatively with secondary users.

Huang \emph{et al.} \cite{SSrelay} investigated the tradeoff between cooperative sensing performance and energy consumption for the scenario shown in Fig. \ref{RFandCoop_fig}(b), where the relays operate with the amplify-and-forward (AF) protocol, and one of the relays will forward the signals of PU to SU in order to help determining the presence and absence of PU. Under given $p_d$ and $p_f$ constraints, the energy consumption for spectrum sensing is minimized over the number of samples (proportional to sensing time) and the amplification gain.

\subsubsection{Resource Allocation for Cooperative Transmission between SUs} With the introduction of the secondary relay, the resource related issues include relay selection, power allocation and subcarrier matching.

Ge and Wang \cite{relayCR1} investigated the power allocation issue for the scenario shown in Fig. \ref{RFandCoop_fig}(a), where multiple PUs co-exist in the three-node AF-relay-enhanced CR network. The optimization problem is formulated to maximize the overall 'bits/Joule' of the OFDM-based CR relaying system under the consideration of many practical limitations, such as transmission power budget of the CR source node (SN) and the relay node (RN), minimal capacity requirement, and interference threshold of the primary users, where the energy consumption includes a constant circuit energy $P_C$.

Similarly, Shaat and Bader \cite{relayJoint4} considered the resource allocation problem in the decode-and-forward (DF) relayed OFDM based CR system. The power allocation is optimized jointly with the subcarrier matching under individual power constraints in the source and relay. The sum rate is maximized while the interference temperature is kept below a pre-specified threshold. Similarly, Zhao \emph{et al.} \cite{5484603} investigated power and channel allocation for a three-node DF relay-enhanced cognitive radio network, where cooperative relay channels are divided into three categories: direct, dual-hop, and relay channels, providing three types of parallel end-to-end transmission.

Chen \emph{et al.} \cite{relayJoint2} studied the distributed relay selection scheme to achieve the goal of maximizing the received reward that is contributed by the system spectrum efficiency and energy consumption. The problem is formulated as a multi-armed restless bandit problem, which has been widely used for stochastic control. The secondary relay node in Fig. \ref{RFandCoop_fig}(a) is equivalent to the arm, in which the state of an arm is characterized by the time-varying channel condition of all related links, spectrum usage and residual energy states. Each arm chooses to be passive or active for relaying the data, and performs the corresponding adaptive modulation and coding (AMC) strategy if the arm is selected. 

Similarly, Luo \emph{et al.} \cite{relayJoint} jointly considered the relay selection scheme and optimal power allocation scheme to achieve a good tradeoff between the achievable data rate and network lifetime. Furthermore, \emph{Chen et al.} \cite{relayJoint3} adopted the multi-armed restless bandit model to improve multimedia transmissions over underlay CR relay networks, while ensuring PUs with a minimum-rate for a certain percentage of time. The cross-layer design approach is used to minimize multimedia distortion, increase spectral efficiency and prolong network lifetime, by jointly considering optimal power allocation, relay selection, adaptive modulation and coding, and intra-refreshing rate.

\subsubsection{Resource Allocation for Cooperative Transmission between PUs and SUs}
With the introduction of cooperative relay in Fig. \ref{RFandCoop_fig}(b), PU engages appropriate SUs to relay its transmission so as to improve primary transmission performance, e.g., enhancing achievable throughput/reliability and/or saving energy. In return, PUs yield a portion of spectrum access opportunities to relaying SUs for secondary transmissions. The strategy of cooperation can be a) {\bf{TDMA based three-phase cooperation}}, i.e., PUs broadcast in the first phase, SUs relay in the second phase, and SUs transmit in the third phase. b) {\bf{FDMA based two phase cooperation}}, in which PU divides its spectrum into two orthogonal subbands, and broadcasts on the first subband in the first phase. SUs relay on the same subband in the second phase, and continuously transmit in both two phases on the second subband.

The work in \cite{cooperativeRelaygame} is concerned with enhancement of spectrum-energy efficiency of a cooperative cognitive
radio network shown in Fig. \ref{RFandCoop_fig}(b). The relay selection and parameter optimization are formulated as a Stackelberg game, which is suitable for problems with a sequential structure of decision-making. Since PUs have higher priority over SUs, PUs are leaders when cooperating with SUs, i.e., PUs decide whether to cooperate and how to cooperate in the first stage. In the second stage, SUs aim to maximize its own utility by playing the non-cooperative power control game.
\begin{figure*}
\includegraphics[width=\textwidth]{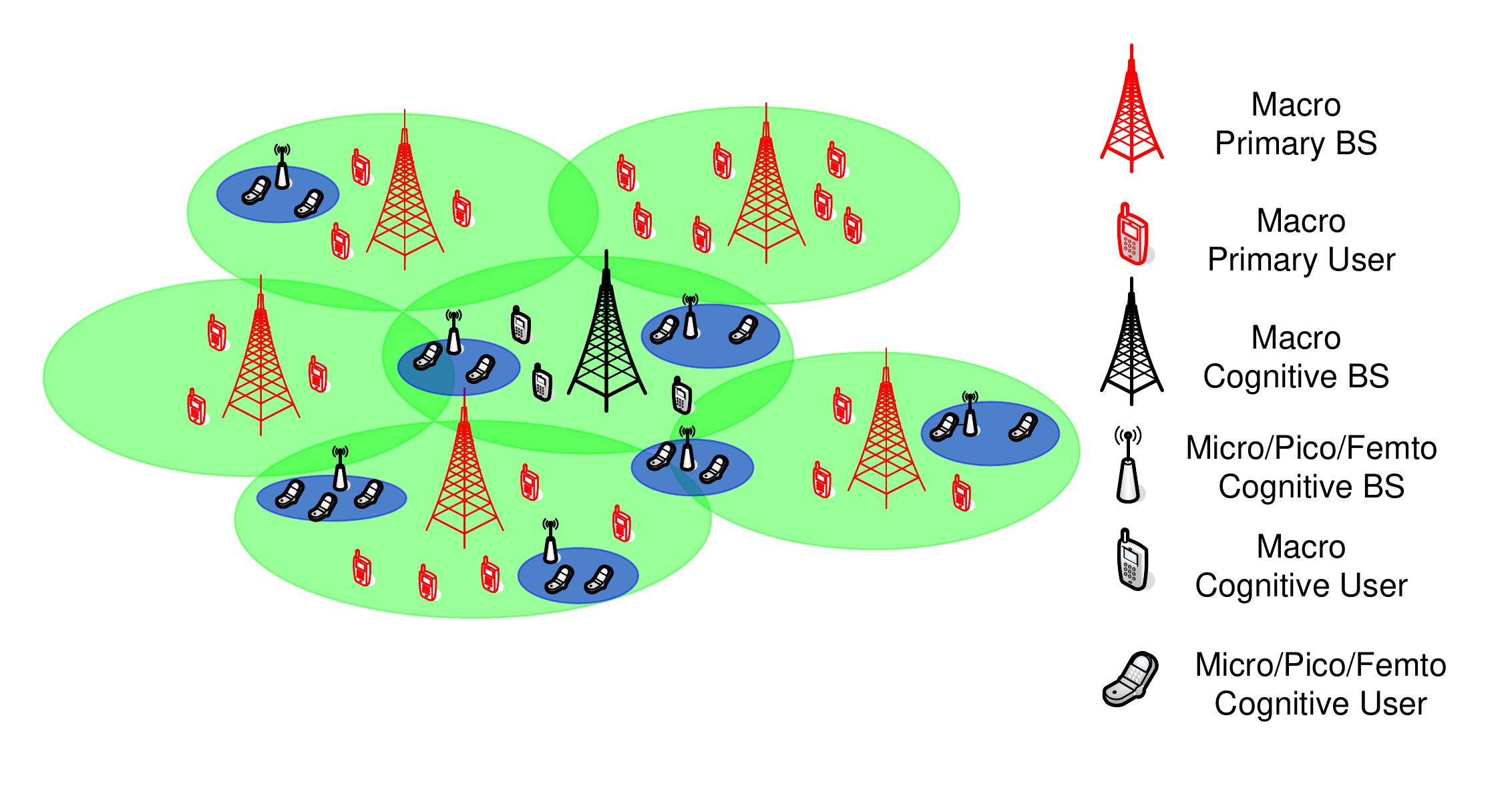}
\caption{Heterogeneous cognitive radio system.}
\label{Heterogeneous_fig}
\end{figure*}

\subsection{Green Cognitive Small Cells}
Although relay and cooperative networks can improve the signal quality of cell edge indoor users to a certain degree, the rapid growth of mobile service usage accelerates the need for novel cellular architectures to meet such demands \cite{HET}. Instead of relying on careful deployment of conventional base stations, the LTE-Advanced or beyond standards propose heterogeneous networks (HetNet's) to resolve the capacity demand issue \cite{Het1}, \cite{Het2}.

HetNet consists of a macrocell network overlayed by small cells. The macro-tier guarantees the coverage, while the overlay network provides means to satisfy the local capacity demand. The small cells in this two-tier architecture can be micro-cells, picocells or femto-cells, where the distinction between them can be found in the size of the cells and the capability of auto-configuration and auto-optimization. 

The reduced cell size can lead to higher spatial frequency reuse and lower power consumption. However, the cross-tier interference will occur if the small cells and the macrocell overlap in their assigned spectrum; the intra-tier interference among multiple small cells will also impair user performance in both cell tiers. Efficient and agile spectrum access enabled by cognitive radio can avoid interference and overcome the coexistence issues in multi-tier networks \cite{coghet}.

As illustrated in Fig. \ref{Heterogeneous_fig}, both macro cell and small cell can be equipped with cognitive functionality. Depending on whether priorities of the secondary users are provisioned, the system could be \emph{open access} or \emph{closed access} \cite{6171997}. When cognitive cells are configured as closed access, only registered secondary users can communicate with their cognitive base stations and primary users can only access their base stations even in a shared subchannel. When cognitive cells are configured as open access, it could be accessed by both its users and all cochannel primary users, and there is no notion of priority for the transmissions of macro BSs \cite{coghet}. To improve the overall energy efficiency which is severely affected by the installation of the additional small cell tier, how to utilize the cognitive capabilities to save energy and mitigate interference is still a research challenge.

\subsubsection{Open access architecture}
The major function of the secondary small cells is to offload traffic from the macro cells. Since the traffic demand highly fluctuates over space, time, and frequency, the small cells do not always have users to serve. By utilizing the spectrum sensing results to determine whether there is traffic to offload, the small cell can decide whether to enter the sleep mode or not. For HetNets with open-access configuration, Wildemeersch \emph{et al.} \cite{cognetEE} evaluated a distributed sleep-mode strategy to minimize the energy consumption of the small cells. Optimal sensing duration and sensing probability are derived under various detection performance constraints and the traffic offloading requirements.

\subsubsection{Closed access architecture}
The transmission of the primary base station has higher priority, and the secondary network has its own users to serve. As compared with the traditional spatial/temporal spectrum sharing scenario, the differences in this case are the confined cell size and transmission power of the cognitive small cells.

Similar to the spatial spectrum sharing scenario in \cite{taoHan}, the secondary cells can grant access to the primary users. In return for the extra power consumption, the secondary cells are guaranteed a certain revenue as the incentive mechanism. Yang \emph{et al.} \cite{6356121} studied the joint power allocation of the macro primary base station and secondary small cells to guarantee the energy-saving and throughput of the primary network. In particular, the utility of the primary users is maximized and the pricing function is incorporated to reflect the virtual money to be paid to the small cells. Xie \emph{et al.} \cite {auctionSmallcell} fully exploited the cognitive capability by considering a wireless network architecture with macro and femto cognitive cells. The 'bits/Hz per Joule' energy-efficient resource allocation problem is modeled as a three-stage Stackelberg game, where the primary network leads the game by offering the selling price in the first stage. The cognitive macro base station will decide whether to buy the spectrum size in the second stage, and allocate the procured spectrum among femtocells and macro secondary users. In Stage III, the femtocell performs power allocation for the femtocell secondary users. A gradient based iteration algorithm is proposed to obtain the Stackelberg equilibrium.

\section{Energy-Harvesting CR}
While all the above techniques optimize and adapt energy usage to achieve energy efficient CR networks, energy related inhibition has not been precluded. For battery powered nodes, they can only operate for a finite duration as long as the battery lasts. For main grid-powered nodes, the primary rationale still pertains to cost efficiency and reduced carbon footprints. As a result, CR has to be optimized to provision green communications by using green power. 

Energy harvesting has been applied to address the problem of tapping energy from readily available ambient sources that are free for users, including wind, solar, biomass, hydro, geothermal, tides, and even radio frequency signals \cite{RF}, \cite{RFHARVESTOR}. Depending on application scenarios, SUs can be sensor nodes in a sensor network, or mobile devices and even base stations in a cellular network; all these devices can be equipped with the energy harvesting capability \cite{Kansal:2007:PME:1274858.1274870}, \cite{Han:2014:PMN}.

The \emph{energy-arrival rate} is a significant metric to evaluate the energy harvesting capability. To guarantee a certain level of stability in energy provisioning, energy sources are normally implemented in a complementary manner. As a result, \emph{passively powered devices} (the energy harvesting generators that do not require any internal power source \cite{RF} and sometimes use different green energy sources) and \emph{hybrid powered devices} (the energy harvesting generators that have a backup non-renewable energy source in case the power provided by energy harvesting is insufficient \cite{Han:2012:OCS}, \cite{Han:2012:ICE}) are attractive as inexhaustible replacements for traditional wireless electronic devices in the cognitive radio networks.

\subsection{Green Energy Harvesting Models}
\begin{figure*}
\includegraphics[width=\textwidth]{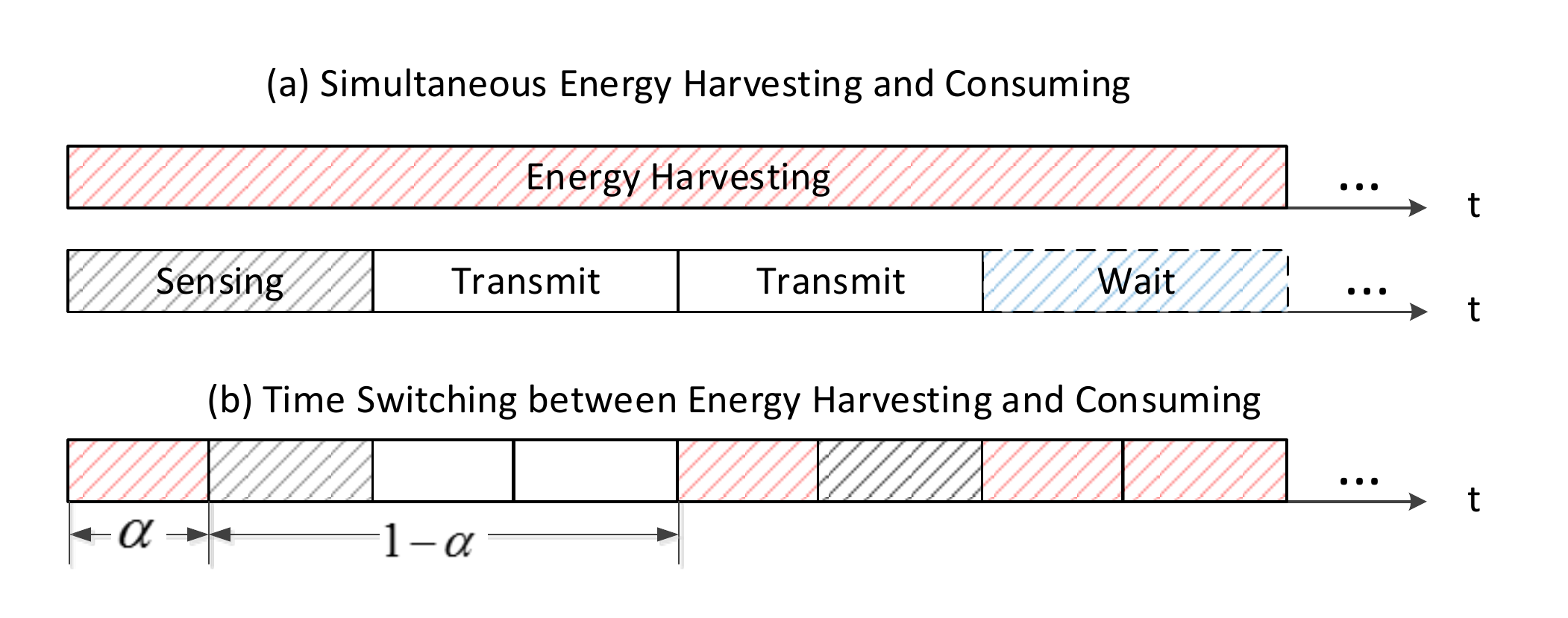}
\caption{The harvest-store-use mechanism of cognitive device.}
\label{harvest}
\end{figure*}

Depending on whether there is a storage capability for the power output of the harvesting system, the generic system model is classified into: 1) the \emph{harvest-use} architecture, which mandates that the instantaneous energy harvesting rate should always be no less than the energy consumption rate \cite{Kansal:2007:PME:1274858.1274870}; 2) the \emph{harvest-store-use} architecture with a storage component (e.g., rechargeable batteries) to hoard the harvested energy for future use. As illustrated in Fig. \ref{harvest}, the energy harvesting process and energy consuming process (i.e., sensing, transmission, reception, etc.) can be scheduled simultaneously, or in a \emph{time switching} way \cite{6489506}. The unified model of the energy harvesting process of these architectures is given in Eq. (\ref{eff1}).
\begin{equation}\label{eff1}
{E_h}={\alpha}\eta {E}
\end{equation}
where ${E_h}$ is the harvested energy, $E$ represents the green energy source, which is normally modeled as a \emph{Markov process} \cite{6710085}. $0 < \alpha\le1$ is the time switching ratio that is used for energy harvesting ($\alpha=1$ for simultaneous energy harvesting and consumption), and $0 < \eta  < 1$ is the energy conversion efficiency.

\begin{figure*}
\includegraphics[width=\textwidth]{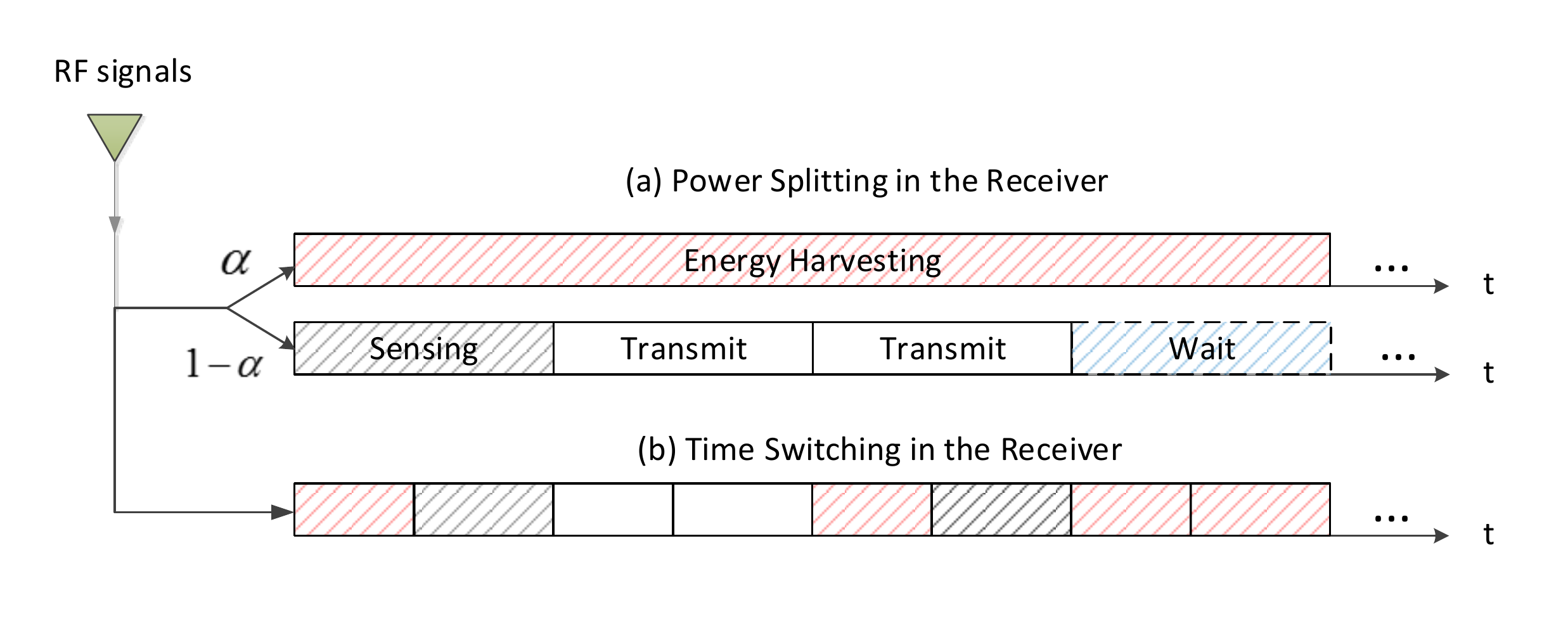}
\caption{The RF energy harvesting mechanism.}
\label{powersplitting}
\end{figure*}

For the green cognitive radio network, RF harvesting is an energy form of particular potential because the green CR can transmit data on the idle spectrum while harvesting energy from the busy spectrum. Another advantage of RF energy is the simultaneous transfer of wireless information and power \cite{6489506}. So, except for the separated energy harvester and information receiver similar to Fig. \ref{harvest}(a), co-located energy harvester and information receiver can adopt two practical architectures: \emph{power splitting} and \emph{time switching}. As depicted in Fig. \ref{powersplitting}(a), when the RF signal reaches the receiver, part of it is used for power extraction and the rest for the simultaneous data detection. Note that Fig. \ref{powersplitting}(b), which requires only one set of antenna for both energy and data, is a special case of Fig. \ref{harvest}(b). The unified model of the RF energy harvesting process of these architectures is given in Eq. (\ref{eff}).

\begin{equation}\label{eff}
{E_h}={\alpha}\eta{\left| h \right|^2}{E}
\end{equation}
where $h$ is the channel condition between the RF energy harvester and the RF energy source. ${\left| h \right|^2}{E}$ represents the energy of the received RF signals. $0 < \alpha\le1$ is the time switching ratio or power splitting ratio that is used for energy harvesting.

For architectures shown in Fig. \ref{harvest}(a) and Fig. \ref{powersplitting}(a), the existing literature assumes that energy harvested in the current time slot can only be used in subsequent time slots, owing to the \emph{energy half-duplex constraint} \cite{6449245}. So, before performing the cognitive functionality, the available residual energy is observable in all of the architectures. However, as compared with the cognitive radio systems powered by traditional on-grid energy, EH CR is different in the sense of the dynamic nature of energy supply, i.e., \emph{opportunistic energy harvesting} makes the energy-arrival rate no longer constant. Accordingly, the power budget is dynamic, namely, \emph{energy causality constraint} (EC-constraint), also referred to as the energy neutrality constraint. The EC-constraint requires that the total consumed energy should not exceed the total harvested energy, which may further be limited by the finite battery capacity \cite{5522465}, \cite{energyCausality}. 

\subsection{Green Energy Utilization and Optimization}
The conventional energy efficient techniques cannot be applied directly into the case where the transmitter or the receiver is subject to intermittent and random harvested energy. For passively powered devices, it is more reasonable to adopt the performance maximization approach: optimize the system function according to harnessed green energy. Although minimizing current energy consumption may bring future reward in performance, the energy minimization approach alone cannot guarantee the system performance, which is the overall design objective. For hybrid powered devices, the three approaches can still be adopted because on-grid non-green power remains part of the energy source. However, energy minimization and utility maximization should factor in the fact that the green energy consumption is free for a given deployed EH system.

\subsubsection{Transmission Policy with Energy Harvester}
For the architecture with separated energy harvester and information transmitter, as illustrated in Fig. \ref{harvest}(a), maximizing the utilization of green energy means tailoring the power management scheme for the random harvested source. Ho and Zhang \cite{6202352} considered the point-to-point wireless data transmission system with the energy harvesting transmitter. They assumed two types of side information (SI) about the harvested energy and time varying channel conditions are available: causal SI (of the past and present slots) or full SI (of the past, present and future slots). Optimal energy allocation is developed to maximize the throughput over a finite horizon. Similarly, optimal energy management schemes are provided to maximize the throughput by a deadline and minimize the transmission completion time of the communication session \cite{5441354}, \cite{5992841}. Also, the throughput optimal policy is proposed for the constrained setting where only online knowledge of the energy arrival process and the channel fade level is available \cite{5992841}.

Furthermore, the energy management policies can be extended to various directions. For example, the works in \cite{6363767} and \cite{6381384} explored the joint source and relay power allocation over time to maximize the maximum throughput of the three node DF relay system, in which both the source and relay nodes transmit with power drawn from energy-harvesting sources. Han and Ansari \cite{Han:2012:ICE} studied the power allocation scheme for the system with hybrid energy supplies. By utilizing the green power thoughtfully, the on-grid power consumption is minimized while the QoS of users in terms of SINR is guaranteed.

Although the energy cannot be consumed before it is harvested, the traffic cannot be delivered before it has arrived. Yang and Ulukus \cite{6094139} proposed the packet transmission policy to minimize the packet transmission completion time in which the transmitter has to wait until enough green energy is harvested in order to adopt a higher transmission rate. For a multi-user system, the multi-user diversity in terms of channel conditions can be explored to enhance the green energy utilization. Therefore, scheduling different users at a given time slot may require different amounts of green energy. By exploring the multi-user diversity, a packet scheduling algorithm may shape BS's energy demands to match the green power generation \cite{6655120}. 


\subsubsection{Reception Policy with Energy Harvester}
To address the opportunistic wireless energy harvesting at the receiver side, the goal is to design the optimal splitting and switching rule for the architectures shown in Fig. \ref{powersplitting}. 

Since the co-channel interference is useful for the wireless RF energy harvester, though it limits the quality of data reception, the time switching rule needs to incorporate the time varying channel condition. To minimize the outage probability or the ergodic capacity for wireless information transfer, while maximizing the average harvested energy, joint optimization of transmit power control of the transmitter, which is plugged with a traditional on-grid power source, and scheduling of information and energy transfer in the receiver, is considered in \cite{6373669}.

The design of the harvesting rule in the receiver is of particular importance to the relay networks because the amount of harvested energy determines the forwarding ability of the relay, while the information received in the relay acts like the data source in the forwarding phase. According to the numerical analysis \cite{6552840}, various system parameters, such as power splitting ratio, energy harvesting time, source transmission rate, noise power, source to relay distance, and energy harvesting efficiency, affect the performance of the three node AF relay channel, where the relay node harvests energy from the received RF signal and uses that harvested energy to forward the source information to the destination.

\subsubsection{Challenges on Green Energy Utilization and Optimization}
Currently, the research on green energy utilization and optimization mainly relies on having the knowledge of the instantaneous energy arrival rate, either assuming that it is already available (offline optimization), or assuming that the causal information or statistical knowledge can be obtained (online optimization) \cite{6710085}.

For the offline optimization, the closed form optimal transmission/reception can normally be obtained, such as variants of the water-filling algorithm \cite{6202352}, \cite{5992841}. Since the assumption of having complete information of the energy arrival rate (past, present and future) is normally unrealistic due to the intermittent nature of the green energy source, various learning algorithms have been studied to estimate the statistical parameters of the energy arrival processes \cite{6477834}. When the energy harvesting profile is modeled as the Markov process, the online energy management scheme is cast under the framework of Markov decision processes (MDPs). However, such decision policies are complex, and thus a simpler alternative is required to analytically balance energy consumption and harvesting \cite{6710085}.
\subsection{Cognitive Functionalities in Energy Harvesting}
\begin{figure*}
\includegraphics[width=\textwidth]{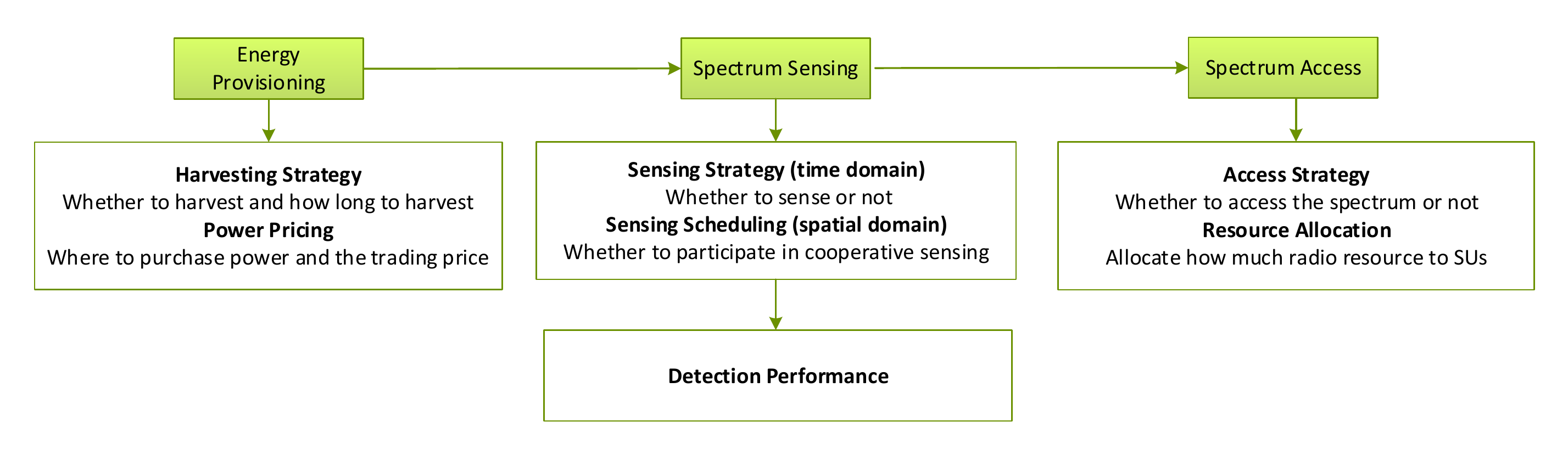}
\caption{Cognitive functionalities in energy harvesting.}
\label{Diagram}
\end{figure*}

Envisioning green energy as an important energy resource, the design and optimization of the cognitive radio network is challenging since the network performance and the user QoS highly depend on the dynamics of the available spectrum and green energy. To alleviate the network from the spectral and energy constraints, the inherent relationships between system performance and the availability of power and spectrum are investigated and illustrated in Fig. \ref{Diagram}.

1) \emph{Harvesting strategy}: In the time domain, \emph{prior} to the sensing strategy and access strategy, harvesting strategy is needed in the time switching architecture to specify whether to harvest and how long to harvest because it affects the EC-constraint, which will influence the subsequent sensing and possible transmission. Yin \emph{et al.} \cite{6654965} studied the harvesting duration and number of channels to be sensed next. The mixed integer non-linear programming (MINLP) problem is formulated to maximize the expected achievable throughput of SU over all of the idle channels in one time slot.

Meanwhile, when considering the architecture of the receiver in Fig. \ref{powersplitting}, how to split the power between RF harvesting and data detection is another research challenge. This is of particular importance for the relaying and cooperative CR networks because the detected data need to be transmitted to the destination by utilizing the harvested energy.

2) \emph{Sensing strategy}: The non-constant EC-constraint will affect the \emph{sensing strategy} because the instantaneous low residual energy may not be able to fully support a complete sensing process, or not be able to complete the data transmission successfully, even the channel will be indicated as idle if SU indeed decides to sense. So, should SU 1) remain idle until the accumulated energy reaches a threshold, which is greater than power needed for a single sensing and transmission, 2) perform sensing and possible transmission when the residual energy is sufficient for a single sensing and transmission, 3) perform sensing and gain spectrum occupancy information when the accumulated energy is above a predefined sensing power threshold, or 4) keep sensing as long as there is energy left? Although the work in \cite{4663925} has already considered the sensing decision policy, the corresponding zero energy arrival rate of the un-rechargeable cognitive device makes the third and forth policy trivial because when the residual energy is less than the power required for a successful sensing and transmission, any sensing strategy afterwards will bring zero payoff. For the cognitive device equipped with energy harvester, however, the fourth policy may be preferable because spectrum sensing not only gains spectrum occupancy information, but also leaves room in the battery for the newly arrived free energy.

3) \emph{Detection performance}: The detection probability and the false alarm probability in Eq. (\ref{sense_f}) should factor in the sensing strategy, owing to the fact that $p_d$ will increase if the cognitive device stays idle without sensing when the channel is busy. On the other hand, when the channel is actually idle, $p_f$ will increase. Modeling the energy arrival process as independent and identically distributed (i.i.d.) sequence of random variables with fixed average rate, Park \emph{et al.} \cite{EHoverlayCR} adopted the sensing decision policy where SU with sufficient energy to carry out spectrum sensing and transmission will sense the channel. The optimal single spectrum detection threshold $\lambda$ is derived to maximize the expected total throughput of SU under energy causality and collision constraints.

4) \emph{Sensing scheduling}: In the spatial domain, the diversity of residual energy among cooperative devices makes the scheduling problem more complex. For instance, to maximize the life time of sensors which are dedicated to perform cooperative spectrum sensing using the harvested energy, an online algorithm is required for the scheduling of each sensor's active time, and the scheme proposed in \cite{eeSSc2} is insufficient since the energy arrival rate is not zero anymore.

When spectrum sensing is shared in a distributed non-cooperative way, existing literature makes the implicit assumption that no explicit relationship exists between the sensing/reporting power and available transmission power, i.e., SU can always transmit with the maximum power. This is reasonable since the steady positive energy arrival rate is guaranteed by the on-grid power source. So, the expected throughput will be a function of sensing policies (whether to contribute to sensing or not), sensing results and channel condition $\gamma$. In the EH-CR environment, however, each SU's utility function should take the available residual energy into account, which will further affect their sensing strategy. In the frequency domain, the secondary system or every SU still has to take the residual energy into consideration in scheduling the sensing process because the subsequent transmission will be affected by it.

5) \emph{Access strategy}: With varying energy arrival rate, the user maybe forced to remain idle on the unoccupied spectrum if the remaining energy is insufficient to perform a successful data transmission, and so the access strategy has to incorporate the dynamic EC-constraint. Park \emph{et al.} \cite{overlayEHCRbourBernoulli} investigated how energy harvesting capability affects the expected total throughput of SU. The optimal sensing and access policy determining which channel to sense and possible access is formulated as a partially observable Markov decision process because the spectrum occupancy state is not fully observable to SU. With a given sensing duration and fixed power required for data transmission in one time slot, a sub-optimal myopic policy, which maximizes the expected immediate throughput of SU is proposed. Furthermore, Park \emph{et al.} \cite{EHrfCRsensingHarvesting} considered a scenario where idle SUs can harvest RF energy from active PUs. The optimal sensing decision policy and access policy are also formulated as POMDP, and a sub-optimal myopic policy is proposed as well.

6) \emph{Resource Allocation}: The harvesting strategy will have an impact on the sensing decision, which further affects the available power for the subsequent data transmission. To achieve better performance of the secondary networks, the dynamic resource allocation mechanism should be able to reveal these internal relationships. Adopting the similar sensing decision policy in \cite{EHoverlayCR}, Chung \emph{et al.} \cite{6692333} studied the power control scheme which maximizes the average throughput during a slot, subject to the collision constraint imposed by PU. Gao \emph{et al.} \cite{PCandSD} jointly considered the sensing strategy and power allocation strategy to maximize the throughput of SU over multiple consecutive time slots. A sub-optimal online algorithm is designed based on the dynamic spectrum state, harvested energy, and the channel fading level.

7) \emph{Power Pricing}: Although current wireless access networks mainly rely on the power grid, which is a large interconnected infrastructure for delivering electricity from power plants to end users, the majority of research endeavors on green cognitive radio has been focused on the locally generated green power, whether by the cognitive device itself, or trade with other green base stations. 

In fact, facing the emerging challenges, e.g., rising energy demand, aging infrastructure, and increasing greenhouse gas emission 
\cite{pp1} -\cite{pp3}, the traditional power grid is becoming smart, which can integrate with renewable green energy resources such as wind and solar power \cite{solarsmart}. The work in {\cite{smartgrid} represents cognitive heterogeneous mobile networks in the smart grid environment. A three-level Stackelberg game is developed to model the problems of electricity price decision, energy efficient power allocation and interference management. Then, a homogeneous Bertrand game with asymmetric costs is used to model price decisions made by the electricity retailers. A backward induction method is used to analyze the proposed Stackelberg game. Simulations are conducted to evaluate the reduction in the operational expenditure and $CO_2$ emissions in cognitive heterogeneous mobile networks.



\section{Green Energy powered Cognitive Radio Networks}
Although we have discussed green powered cognitive devices in the previous sections, we try to further improve the energy efficiency by utilizing the green power in the cognitive radio networks. As the smart grid advances and develops, green power farms that harvest energy from green sources, e.g., solar energy and wind energy, can substantially reduce carbon footprints. Also, the penetration of distributed green energy capitalization is galvanizing worldwide. In the near future, consumers are allowed to contribute their clean energy back to the grid \cite{NIRWAN}.

Taking advantages of distributed electricity generators, telecommunications equipment manufacturers, such as Nokia Siemens and Ericsson, have designed and built green energy powered off-grid base stations to reduce OPEX of mobile networks in rural areas \cite{Han:2014:PMN}. The power consumption of a mobile network depends on the energy generated from the ambient source, the storage capacity of the battery, the traffic demands and the wireless channel conditions, which are all dynamic processes. In a cognitive radio based wireless network, the operation of the system even depends on the fluctuation of the available spectrum, as shown in Fig. \ref{model_fig}.

\begin{figure}
\centering
\includegraphics[width=3in]{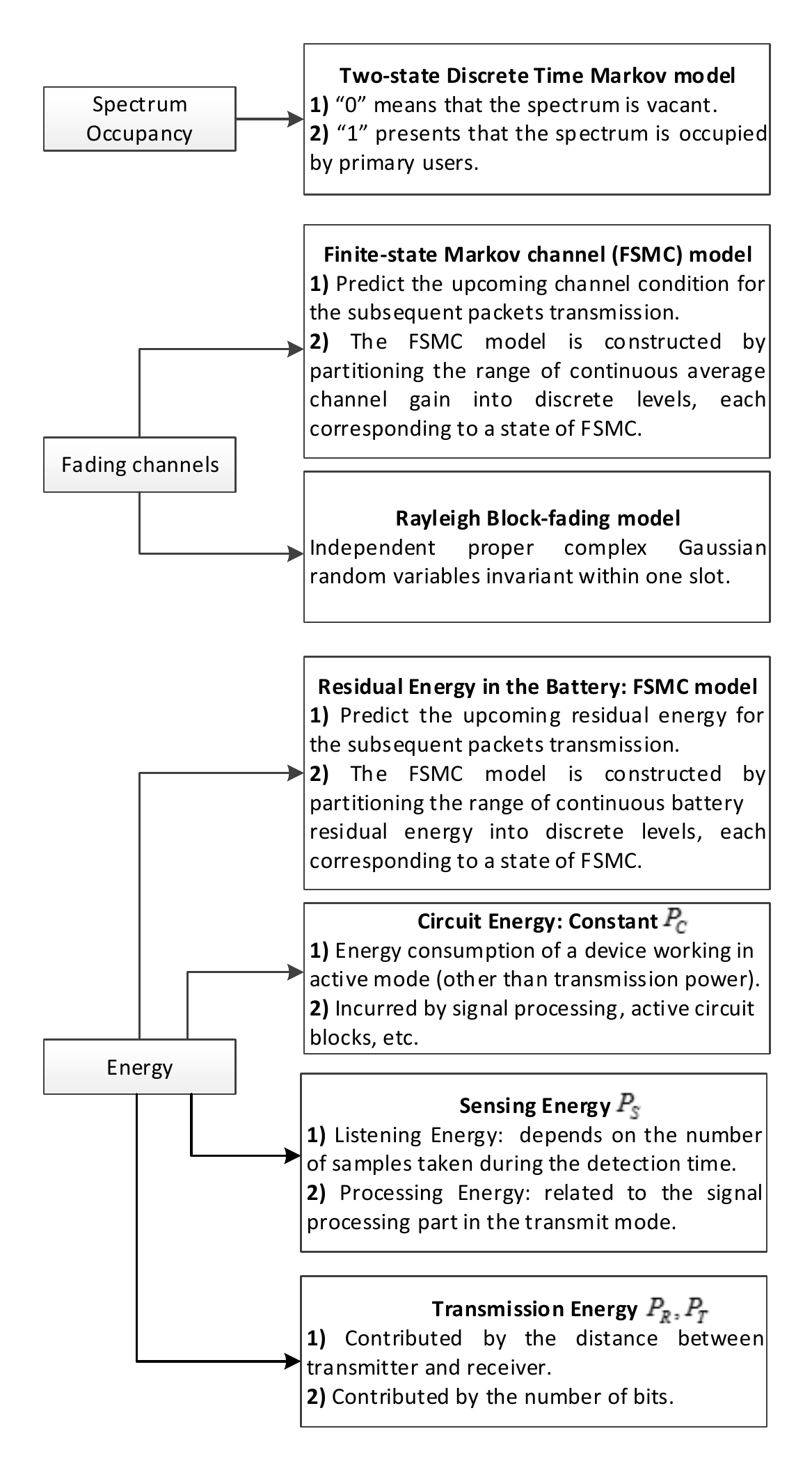}
\caption{Common Models used in CR network.}
\label{model_fig}
\end{figure}

In addition to the utilization of standalone power generators, optimizing green energy enabled mobile networks involves power sharing/trading with other distributed green generators and power purchasing from the green power farms of the smart grid, as illustrated in Fig.\ref{grid}.

\begin{figure*}
\centering
\includegraphics[width=\textwidth]{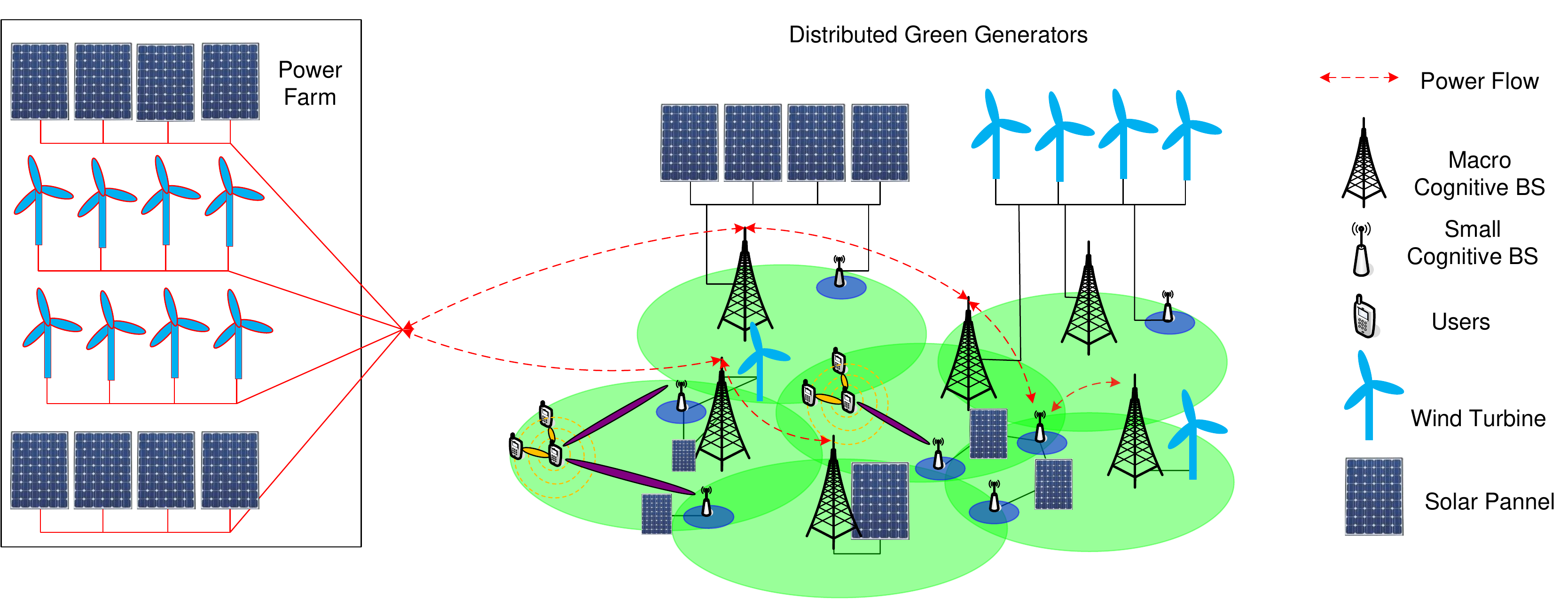}
\caption{Green energy powered cognitive radio network.}
\label{grid}
\end{figure*}

\subsection{Wireless Network Powered by Distributed Green Generators}
Taking advantages of the characteristics of wireless communications, the power sharing with neighboring cells can be facilitated through traffic offloading. Various user-BS association schemes or cell size adapting schemes can realize this implicit power sharing method. Han and Ansari \cite{aaaa} considered the centralized user association scheme by taking into account of the green energy storage and traffic queue of multiple base stations, a green energy aware and latency aware scheme is proposed to minimize the sum of the weighted traffic delivery latency of BSs in a heterogeneous mobile network. Meanwhile, the user can associate with multiple base stations, as in coordinated multipoint transmission (CoMP), such that the diversity of channel conditions, traffic load and amount residual green energy can be utilized jointly for the data transmission. Similarly, for the cell size adapting scheme, the BSs can decide whether to expand and cover more active users, or shrink smaller until entering the sleep state.


With the grid infrastructure, the power generated by the distributed green generators can be directly shared. For instance, when users with traffic are located far away from the base stations with ample energy storage, traffic offloading is not power economical or even feasible due to the fading of the wireless channel. In this case, it is more rational to transmit the energy through the power line directly. Furthermore, power trading can be implemented where the node with traffic to serve and the node with ample power storage can be selfish users. The trading process can also take the form of traffic offloading, power transmitting, or both.

\subsection{Wireless Network Powered by Green Power Farms}
Although power can be shared/traded as a fundamental resource element which determines the performance of the wireless system, there is a side effect from power and interference. Less power means less interference to the neighbors. Accordingly, the neighbors' power will be consumed less. Kwak \emph{et al.} \cite{6319333} implemented the centralized power sharing scheme through interference management. For a given power budget which can be used to purchase a certain amount of power from the electricity supplier, or for a given capacity of the green power farm, how to allocate the power into multiple base stations and time slots determines the long-term utility function of the network. The utility based on SINR is optimized jointly by user scheduling and BS power control for different power sharing constraints in spatial and time domains.

When smart grid with green power plants is the power provider for the wireless cellular network, the energy price has a great impact on the energy efficiency in the cellular network. Owing to the uncontrollable nature of the green source, the generated green power may be surplus, and this may harm the energy suppliers since energy storage cost will impact the revenue \cite{5558703}. As a result, balancing the market through pricing is extremely important in the green environment. 

Bu \emph{et al.} \cite{6210335} considered a system with multiple electricity retailers, and multiple clusters of BSs. BSs in one cluster can go to sleep or be active and participate in the CoMP transmission with other active cluster members. Considering the pollutant level of and the price offered by each retailer, independent clusters need to decide which retailers to procure electricity from and how much electricity to procure. The system has been modeled as a two-level Stackelberg game, where 1) the Stackelberg leaders, the electricity retailers, try to maximize the revenue; 2) the Stackelberg follower, the cellular network, tries to maximize the utility function with regard to blocking probability, power cost and pollutant levels.

\subsection{Cognitive Radio Network Powered by Green Energy}
The green energy enabled wireless network can be provisioned by a green power farm or distributed green generators. Table \ref{tab_1} describes the pros and cons of these two types of provisioning; in particular, the green power farm is costlier to build while the energy provisioned by distributed green generators is less stable.

Whether being powered by a green power farm or distributed green generators, cognitive radio network imposes more challenges for green energy utilization because the nodes with available spectrum, the nodes with sufficient power and the ones with data traffic to serve may all be different. Emerging methods which can balance the spectrum, power and traffic within the secondary network are desirable and briefly discussed next.

1) Sharing/Trading sensing results is a form of power balancing. For instance, traffic arrives at a SU transmitter, which is supposed to perform spectrum sensing and data transmission to its intended SU receiver. A third CR node near the SU receiver may keep sensing and offer the sensing results if the aforementioned SU transmitter is in power shortage. With a sufficient large amount of power, this CR node can even directly offload the data traffic from the SU transmitter. In this way, the transmission power will be saved due to a shorter distance between the third CR node and the SU receiver. Mutually, if all of the cognitive radio devices act like each other's remote backup power storage, the temporal and spatial power sharing within the networks is realized.

2) Mobile charging which relies on the wireless energy transfer can be another approach to balance the power within a cognitive radio network. Mobile vehicles/robots, which carry high volume batteries \cite{6502505}, can serve as back up mobile power storage and periodically deliver energy to cognitive devices with insufficient energy supply.

\begin{table*}[t]
\vspace{-3em}
\caption{Green energy enabled wireless network} 
\centering\  
\begin{tabular}{|c|c|c|} 
\hline                        
Energy generator&	Green power farm	&Distributed green generators\\\hline
Scale&	Large and remote energy supplier	&Small and local energy suppliers\\\hline
Energy storage unit	&High capacity	&Low capacity\\\hline
Energy availability	&More stable	&Less stable\\\hline
Energy transfer method&	Grid structure	&Directly connected to the BSs\\\hline
Energy sharing within the network	&\multicolumn{2}{|c|}{Grid structure or traffic offloading}	\\\hline
\end{tabular}
\label{tab_1}
\end{table*}

3) Power balancing can rely on the grid architecture to transmit power, especially when sufficient power source and cognitive devices in need of power are separated far away. Since spectrum opportunity is a local concept, offloading the sensing tasks or trading sensing results will be unreliable. Also, moving vehicles in long distance will result in high energy consumption.


%
%
%

\section{Conclusion}
This article surveys various developments and advances on energy efficient cognitive radio functionality from spectrum sensing and analysis, spectrum management and handoff, to spectrum sharing and allocation. The state of the art of the energy efficient CR based wireless access network, such as relay and cooperative radio and small cells, is reviewed from each aspect. Recent advances and challenges in research related to energy harvesting based green cognitive radios are elicited. Yet, many challenging problems remain to be tackled. This article will hopefully help readers to jump start further research in provisioning green energy powered cognitive radio networks.
\bibliographystyle{IEEEtran}
\bibliography{mybib}

\end{document}